\documentclass[cmbright]{cmcma}

\usepackage{subfigure}
\usepackage[linesnumbered,ruled]{algorithm2e}

\usepackage{moreverb}

\usepackage[colorlinks,bookmarksopen,bookmarksnumbered,citecolor=red,urlcolor=red]{hyperref}

\newcommand\BibTeX{{\rmfamily B\kern-.05em \textsc{i\kern-.025em b}\kern-.08em
T\kern-.1667em\lower.7ex\hbox{E}\kern-.125emX}}

\def\volumeyear{2022}

\begin{document}

\setcounter{page}{1}

\runninghead{F. Salehi et al.}

\title{Numerical investigation of Differential Biological Models via  RBF collocation Method with Genetic Strategy }

\author{Fardin Salehi 1$^a$, Soleiman Hashemi-Shahraki 2$^{a}$ and  Mohammad Kazem Fallah 3$^a$ and Mohammad Hemami 4$^{a, b}$}

\address{$^a$ Department of Computer and Data Sciences, Faculty of Mathematical Sciences, Shahid Beheshti University, Tehran, Iran,
\\
$^b$ Department of Cognitive Modeling, Institute for Cognitive and Brain Sciences, Shahid Beheshti University, Tehran, Iran.}

\corraddr{Mohammad Hemami}

\begin{abstract}
In this paper, we use radial basis function collocation method for solving the system of differential equations in the area of biology. One of the challenges in RBF method is picking out an optimal value for shape parameter in Radial basis function to achieve the best result of the method because there are not any available analytical approaches  for obtaining optimal shape parameter. For this reason, we design a genetic algorithm to detect a close optimal shape parameter. The experimental results show that this strategy is efficient in the systems of differential models in biology such as HIV and Influenza. Furthermore, we show that using our pseudo-combination formula for crossover in genetic strategy leads to convergence in the nearly best selection of shape parameter. 
\end{abstract}

\MOS{65L05; 65L60}

\keywords{Radial basis function; Genetic algorithm; HIV; Influenza; Shape parameter}

\maketitle

\vspace{-6pt}
\section{Introduction}\label{A}
In the past few years, various mathematical models have been used to investigate biological and clinical problems. Various examples of mathematical models have been presented hitherto, including HIV\cite{perelson1989modeling,perelson1999mathematical}, Influenza\cite{kermack1927contribution}, Covid-19\cite{vespignani2020modelling,samui2020mathematical} and Ebola\cite{berge2017simple,agusto2017mathematical}. In this paper, we will focus specifically on the simulation of HIV and Influenza models using the combination of RBF method and genetic algorithm and we show that this method is very efficient and robust for simulating biological models.
\subsection{HIV Infection CD4+T Cells}
Acquired Immune Deficiency Syndrome (AIDS), first-time appeared in the continent of America  in 1981\cite{bennett2014principles}. The human immunodeficiency Virus (HIV) in short HIV is the cause of the illness that attacks vital cells such as Dendrite cells, helper lymphocyte particularly   CD4+T cells and infects them and gradually,  the immune system will be destroyed\cite{cunningham2010manipulation}. 
This process may take from 6 months to 10 years. The mathematical model of HIV-infected CD4+T cells described by Perelson and Nelson in 1991\cite{perelson1989modeling,perelson1999mathematical}. HIV model investigates the concentration of susceptible CD4+T cells infected by the HIV viruses. It is obvious that presenting a mathematical model is an easier study of the behavior of the system and helps  the process of detecting or improving disease.
Let $\textbf{T}(t)$ be the concentration of susceptible CD4+T cells, $\textbf{I}(t)$ be CD4+T cells infected by the HIV virus and $\textbf{V}(t)$ be free HIV particular in the blood at the time. Thus, the mathematical model of the HIV-infected CD4+T cell on a couple system of the ordinary differential equation will be presented as follows: 
\begin{eqnarray}
\begin{cases}
&\dfrac{d}{dt}\textbf{T}(t)=s-\alpha \textbf{T}(t)+r\textbf{T}(t)(1-\dfrac{\textbf{T}(t)+\textbf{I}(t)}{T_{max}})-k\textbf{V}(t)\textbf{T}(t),~~~~ \textbf{T}(0)=T_0,\\
&\dfrac{d}{dt}\textbf{I}(t)=k\textbf{V}(t)\textbf{T}(t)-\beta\textbf{ I}(t),~~~~~~~~~~~~~~~~~~~~~~~
~~~~~~~~~~~\quad\quad\quad\quad\textbf{I}(0)=I_0,~~~
0\leq t\leq R \leq \infty \\
&\dfrac{d}{dt}\textbf{V}(t)=N\beta \textbf{I}(t)-\gamma \textbf{V}(t),~~~~~~~~~~~~~~~~~~~~~~~
~~~~~~~~~~~~~~~~~\quad\quad\quad\textbf{V}(0)=V_0,
\end{cases}
\label{HIVMODEL}
\end{eqnarray}
where $R$ is a positive constant and other parameters have been shown in Table \ref{Tab1}.
\begin{table}[htbp!]
\caption{Parameters in HIV infected CD4+T cells model}
\label{Tab1}
\centering \begin{tabular}{|l|l|}
\hline 
$\alpha$ & Natural turnover  rates of uninfected T cells \\ 
$\beta$ & Infected T cells \\ 
$\gamma$ & Virus particles \\ 
$k$ & Infection rate \\ 
$s$ & Rate of constructing T cells  \\ 
$r$ & Rate of T cells mitoses \\ 
$N$ & Virus particle from each infected T cell \\ 
$T_{max}$ & Maximum T cell concentration in the body \\ 
\hline 
\end{tabular} 
\end{table}

It is noteworthy that there is no exact solution for HIV model. Ergo, the numerical methods are used to solve it. Table \ref{Tab2} shows some approaches applied to this model.

\begin{table}[hbpt!]
\caption{Used technique for solving HIV model}
\label{Tab2}
\centering \begin{tabular}{|l|l|c|}
\hline
Author(s)&Method&Year\\
\hline 
Merdan\cite{merdan2007homotopy} & Homotopy Perturbation method(HPM) & 2007 \\ 
Alomari et al.\cite{ghoreishi2011application} & Homotopy Analysis method(HAM) & 2011 \\ 
Merdan et al. \cite{merdan2011approximate} & Variational Iteration Method(VIM) & 2011 \\ 
Ongun\cite{ongun2011laplace} & Laplace Adomian Decomposition Method(LADM) & 2011 \\ 
Do\u{g}un\cite{dougan2012numerical} & Multistep Laplace Adomian Decomposition Method(MLADM) & 2012 \\ 
Khan et al.\cite{khan2013efficient} & Iterative Homotopy Perturbation Transform Method(IHPTM) & 2012 \\ 
Y\"{u}zba\c{s}i\cite{yuzbacsi2012numerical} & Bessel Collocation Method(BCM) & 2012 \\ 
Atangana et al.\cite{atangana2014computational} & Homotopy Decomposition Method(HDM) & 2014 \\ 
Chen\cite{chen2015adomian} & Pad\'{e}-Adomian Decomposition Method(PADM) & 2015 \\ 
Venkatesh et al.\cite{venkatesh2016new} & Legendre Wavelets Method(LWM) & 2016 \\ 
Kajani et al.\cite{gandomani2016numerical} & M\"{u}ntz-Legendre Method(MLM) & 2016 \\ 
El-Baghdady et al.\cite{el2017spectral} & Legendre Collocation Method(LCM) & 2017\\
Parand et al.\cite{parand2018shifted} & Shifted Lagrangian Jacobi Method (SLG) & 2018\\
Parand et al.\cite{parand2018quasilinearization} & Quasilinearization-Lagrangian Method (QLM) & 2018\\
Parand et al.\cite{parand2018pseudospectral} & Pseudospectral Legendre Method (PLM) & 2018\\
Parand et al.\cite{parand2018shifted} & Shifted Boubaker Lagrangian Method (SBLM) & 2018\\
Parand et al.\cite{parand2019numerical} & Shifted Chebyshev Polynomial Method (SCP) & 2019\\
Umar et al.\cite{umar2020stochastic} & Genetic Algorithm Active Set Method (GA-ASM) & 2020\\
Oluwaseun et al.\cite{oluwaseun2021block} & Block Method (BM) & 2021\\
Thirumalai et al.\cite{thirumalai2021solution} & Spectral Collocation Method (SCM) & 2021\\
Umar et al.\cite{umar2021neuro} & Neuro Swarm Intelligent Computing (NSIC) & 2021\\
Hassani et al.\cite{hassani2022optimal} & Generalized Shifted Jacobi Polynomials (GSJP) & 2022\\
\hline 
\end{tabular} 
\end{table}
\subsection{Influenza}
Influenza virus causes a type of disease named {\it Influenza} or {\it Flu} that is divided into four classes A, B, C and D\cite{longo2012harrison}. From the perspective of the epidemic, class A is the most significant class. That is because this type is able to merge and rebuild its genes with host gene\cite{anderson1992infectious,webster1992evolution}. The mathematical model of Susceptible-Infected-Removed (SIRC) for displaying the outbreaks of Influenza in population is defined by Kermack and McKendrick \cite{kermack1927contribution}. This is a system of the  differential equation as follows: 

\begin{eqnarray}
\label{fluI}
\begin{cases}
&\dfrac{d}{dt} \textbf{S}(t)=\mu (1-\textbf{S}(t))-\beta \textbf{S}(t)\textbf{I}(t)-\gamma \textbf{C}(t)~~~~~~~~~~~\quad\textbf{S}(0)=S_0,\\
&\dfrac{d}{dt}\textbf{I}(t) =\beta \textbf{S}(t)\textbf{I}(t) +\sigma \beta \textbf{C}(t)\textbf{I}(t) -(\mu +\theta)\textbf{I}(t),~~~~~~~\textbf{I}(0)=I_0,\\
&\dfrac{d}{dt} \textbf{R}(t)= (1-\sigma)\beta \textbf{C}(t)\textbf{I}(t)+\theta \textbf{I}(t) -(\mu + \delta)\textbf{R}(t),~~~\textbf{R}(0)=R_0,\\
&\dfrac{d}{dt} \textbf{C}(t)= \delta \textbf{R}(t)- \beta \textbf{C}(t)\textbf{I}(t)-(\mu+\gamma)\textbf{C}(t)~~~~~~~~~~\quad\textbf{C}(0)=C_0.
\end{cases}
\label{SIRCMODEL}
\end{eqnarray}
where $\textbf{S}(t), \textbf{I}(t), \textbf{R}(t)$ and $\textbf{C}(t)$  mean ratio susceptible, infections, recovered and cross immune respectively. Other parameters are shown in Table \ref{Tab3i}.
\begin{table}[htbp!]
\caption{Parameters in Flu model}
\label{Tab3i}
\centering \begin{tabular}{|l|l|}
\hline 
$\mu$ & Mortality rate  \\ 
$\theta$ & Improve infection each year\\ 
$\delta$ & Progression from recovered to cross-immune each year  \\ 
$\gamma$ &Progression from recovered to susceptible each yer \\ 
$\sigma$ & Rate of cross-immune into the infective   \\ 
$\beta$ & Contact rate  \\ 
\hline 
\end{tabular} 
\end{table}
This model studied by Khader et al.\cite{khader2014numerical} using Chebyshev spectral method in 2014. Table \ref{Tab3} shows the applying methods to SIRC model.
\begin{table}[hbpt!]
\caption{Used techniques for solving SIRC model}
\label{Tab3}
\centering \begin{tabular}{|l|l|c|}
\hline
Author(s)&Method&Year\\
\hline 
El-Shahed et al.\cite{el2011fractional} & Non-standard Finite Difference(NSFDM) & 2012 \\
Ibrahim et al. \cite{ibrahim2013new} & Modified differential transform method(MDTM) & 2013\\
Zeb et al.\cite{zeb2013analytic} & Multi-step generalized differential transform method(MGDTM) & 2013\\ 
Khader et al.\cite{khader2014numerical} & Chebyshev spectral method(CSM) & 2014 \\ 
Khader et al.\cite{khader2014legendre} & Legendre spectral method(LSM) & 2014\\
Gonz\'{a}lez-parra et al.\cite{gonzalez2014fractional} &  Gr\"{u}nwald –Letnikov method(GLM) & 2014\\
\hline 
\end{tabular} 
\end{table}

\subsection{Meshfree Method}
Firstly, the Meshfree methods introduced by Monaghan and Gingold in 1977. They enlarged a Lagrangian method according to Kernel estimate method\cite{gingold1977smoothed}. A number of meshfree methods such as smoothing particle hydrodynamic (SPH)\cite{takeda1994numerical,colagrossi2003numerical}, Element-Free Galerkin (EFG)\cite{krysl1999element,belytschko1995element}, Reproducing Kernel method (RKM)\cite{abbasbandy2015shooting,azarnavid2015picard}, Meshless local Petrov-Galerkin (MLPG)\cite{rad2015pricing,atluri1998new}, Comapctly supported radial basis function method(CSRBF)\cite{hemami2021phase,khalili2022local}, Radial basis function finite difference method(RBFFD)\cite{moayeri2022npds,hemami2020use}, Radial basis function differential quadrature method(RBF-DQ)\cite{parand2017two,parand2021unsteady} and Kansa method (KM)\cite{parand2013kansa,sharan1997application} are used for solving differential equations (DEs). The appearance of meshfree methods was through the difficulty of the classic methods such as Finite Element method (FEM)\cite{taylor1973numerical,johnson2012numerical} and Finite Difference method (FDM)\cite{dehghan2006finite,dehghan2004weighted} which require a mesh of points for solving problems. In these methods, rising problem dimensions causes increasing complexity (the order of construction of the mesh); furthermore, in meshfree we have no need to make any grid, and scattered points are used instead. 
RBF method as a meshfree approach utilizes as Trial functions of kind (Global/Compact support) Radial basis functions (RBFs)Table \ref{tab:1} demonstrates the RBF types. The main advantages of RBF method are the simplicity, high accuracy, and capability of being applicable in high dimension problems. In addition to these advantages, there exist two main challenges that all methods based on RBFs are faced with; selecting Shape parameters (SP) and distribution of collocation points. Choosing an inappropriate SP decreases the performance of method or even it will be unusable when the method is ill-conditioned. It seems that amount of optimal Shape parameter (oSP) depends on equation state, dimension and etc. Thus, any comprehensive formula not found hitherto for recognizing optimal SP in RBFs. Instead of choosing a proper SP, Many researchers offered different formulas; however, these formulas are applicable only in some special cases.  In \cite{liu2005introduction} SP decomposed to a dimensionless size of support domain $(\alpha_s)$ and a nodal spacing near the point at the center $(d_c)$, where $c=\alpha_sd_c$. Hardy \cite{hardy1971multiquadric} suggested using (inverse) multiquadric formula as follows:
$$c=\frac{1}{0.815\varepsilon}$$
where $\varepsilon=\frac{1}{N}\Sigma_{i=1}^{N}\varepsilon_i$, and $\varepsilon_i$ is the distance of the center from its closest neighbor. Rippa\cite{rippa1999algorithm} used the Predictive residual sum of square (PRESS)  algorithm for calculating a proper SP. Leave on-out cross validation (LOOCV) approach  \cite{fasshauer2007choosing}and Craven and Wahba \cite{craven1978smoothing} which emanated in the statistics literature used for finding optimal SP. Esmaeilbeigi et al. \cite{afiatdoust2015optimal,esmaeilbeigi2014new} employed the genetic package of MATLAB for solving a number of DEs. The following formula is proposed in \cite{kansa1990multiquadrics,sarra2005adaptive} to calculate a reasonable SP
$$c_i=\sqrt{c^2_\alpha(\frac{c^2_\beta}{c^2_\alpha})^\frac{i-1}{n-1}}$$
where $n$ is the number of points and $c_\alpha$ is the smallest and $c_\beta$ is the biggest selected parameter in the domain of candidate SPs. Similarly, in \cite{sarra2005adaptive,xiang2012trigonometric} the SP is obtained by 
$$c_i=c_\alpha +(c_\beta- c_\alpha)\Delta_{rand},$$
where $\Delta_{rand}$ is a random number in arbitrary domain. 

In this paper, we suggested a Meta-heuristic continues Genetic algorithm (CGA) choose a near optimal SP, based on the average of summation of the residual 2-norm (ASN2R) and the average of summation of the relative error (ARE)  for the solution of differential equation systems in Biology sciences.

\begin{table} 
\caption{Some radial basis functions$~_c\psi$, ($r=\|x-x_i\|=r_i$), $c>0$}
\label{tab:1} 
\centering \begin{tabular}{lll}
\hline\noalign{\smallskip}
Class & Name of function & Definition \\
\noalign{\smallskip}\hline\noalign{\smallskip}
1 & Multiquadrics(MQ) & $\sqrt{r^2+c^2}$ \\
& Inverse Multiquadrics (IMQ) & $\frac{1}{\sqrt{r^2+c^2}}$ \\
& Gaussian (GA) & $\exp(-c^2r^2)$ \\
& Inverse Quadrics (IQ) & $\frac{1}{r^2+c^2}$\\
& Hyperbolic Secant (sech) & $\sec h(c\sqrt{r})$\\
\hline
2 & Thin Plate Spline (TPS) & $(-1)^{k+1}r^{2k}\log(r)$\\
& Conical Spline & $r^{2k+1}$\\
\hline
3 & Wendland$_{3,0}$ & $(1-r)^2_+$\\
& Wu$_{3,3}$ & $(1-r)^4_+(16+29r+20r^2+5r^3)$\\
&Oscillator$_{1,3}$ & $(1-r)^4_+(1+4r-15r^2)$\\
&Buhman$_1$ & $12r^4 \log r-21r^4+32r^3-12r^2+1$\\
\hline 
4 & Platte$_{a,b,c}$ & $\cos(cr)\exp(\frac{-b}{(1-r^2)^a}+b)$\\
\noalign{\smallskip}\hline
\end{tabular}
\end{table}

\subsection{Genetic Algorithm}
Genetic algorithm (GA) is a search and optimization approach based on the Genetic principles and natural selection. A GA starts with processing a population of candidate solutions (called individuals or chromosomes) with different competencies. During this process (called evolution), GA changes the population and generates some solutions close to optimal competency (maximum benefit or minimum cost). John Holland invented original GA  in the early 1970s\cite{holland1992adaptation}. He also proposed a theoretical basis for GA according to the Type theory. In the following, David E Goldberg\cite{goldberg1991real} extended GA concept and applied it to encode and solve different problems in miscellaneous fields. GA has many advantages over other optimization methods like:
\begin{itemize}
\item Practicable on both discrete and continuous data, 
\item No need to derivative of objective function (fitness function),
\item Usable in multivariate functions,
\item High potential for parallelization, 
\item Calculating a set of appropriate (close to optimal) solutions,
\item Expandable on experimental, analytical and numerical data.
\end{itemize}

 The main objective of a meta-heuristic algorithm is finding a close-minimum to global minimum (maximum) solution by escaping from local minimum (maximum) solutions. 
Universally, GA is classified to DGA (Discrete GA) and CGA (Continuous GA). In this article, we use CGA  to find a close to optimal SP ($\epsilon_\varrho$-optimal Shape parameter) around a specified interval in the RBF method, where $\varrho$ is either ASRN2 or ARE strategies. 

\section{Methodology}
\subsection{RBF approximation}
Let $\psi:\mathbb{R}^+\rightarrow \mathbb{R}$ be a continuous function with $_c\psi(0)\geq 0$. A radial basis function on $\mathbb{R}^d$ is a function of the  form 
$$_c\psi(\|x-x_i\|),$$
where $x$,$x_i \in \mathbb{R}^d$ and $\|.\|$ denote the Euclidean distance between $x$ and $x_i$s. By choosing $N$ points $\{x_i\}_{i=1}^N$ in $\mathbb{R}^d$ and by defining 
$$s(x)=\sum_{i=1}^N\xi_i  ~_c\psi(\|x-x_i\|);~~\xi_i\in \mathbb{R} ,$$
where $s(x)$ is called a radial basis functions mesh\cite{wendland2004scattered,fasshauer2007meshfree}.
To approximate one-dimensional function $f(x)$, we can illustrate it with an RBF as 
\begin{equation}
f(x)\approx f_n(x)=\sum _{i=1}^N \xi_i ~_c\psi_i(x)=\vec {_c\Psi}^T (x)\vec\Xi,
\label{approximation}
\end{equation}
in which, 
\begin{eqnarray}
&&_c\psi_i(x)=~_c\psi(\|x-x_i\|),\nonumber\\
&&\vec{_c\Psi}^T(x)=[~_c\psi_1(x), ~_c\psi_2(x), \cdots, ~_c\psi_N(x)],\nonumber\\
&&\vec\Xi = [\xi_1, \xi_2, \cdots, \xi_N],\nonumber
\end{eqnarray} 
$x$ is the input and $\xi_i$s are the collection of coefficients to be determined.
By selecting $N$ points ($x_j, j=1, 2, \cdots, N$) in interval: 
$$f_j(x)=\vec{_c\Psi}^T(x_j)\vec\Xi$$
To sum up the discussion of the coefficients matrix, we define 
\begin{equation}
\label{system1}
\mathop{\textbf{M}}\limits^{\smallfrown}\vec\Xi =\vec F,
\end{equation}
where 
\begin{eqnarray}
&&\vec F=[f_1, f_2, \cdots, f_N]^T\nonumber\\
&&\mathop{\textbf{M}}\limits^{\smallfrown}= [\vec{_c\Psi}^T(x_1), \vec{_c\Psi}^T(x_2), \cdots, \vec{_c\Psi}^T(x_N)]^T\nonumber\\
\end{eqnarray}
By solving the system(\ref{system1}), the unknown coefficients $\vec\Xi$ will be attained. 
\subsection{Solving models by RBF method}
Both models are system of first-order differential equations, so we define the solution functions and it's first-order derivatives as follows:
\begin{eqnarray}
U(t) \simeq U_N(t) = \sum_{i=0}^N  a\#_i~_c\psi_i(t),\\
\frac{dU(t)}{dt} \simeq \frac{dU_N(t)}{dt} = \sum_{i=0}^N a\#_i ~_c\psi'_i(t).\label{eqder}
\end{eqnarray}
where $~_c\psi$ is RBF. In addition, solution function $U(t)$ and unknown coefficient $a\#$ are also defined separately according to the unknown functions of the models as follows
\begin{eqnarray}
T(t) \simeq T_n(t) = \sum_{i=0}^n  a1_i~_c\psi_i(t),\\
I(t) \simeq I_n(t) = \sum_{i=0}^n  a2_i~_c\psi_i(t),\\
V(t) \simeq V_n(t) = \sum_{i=0}^n  a3_i~_c\psi_i(t),
\end{eqnarray}
for HIV and 
\begin{eqnarray}
S(t) \simeq S_n(t) = \sum_{i=0}^n  a1_i~_c\psi_i(t),\\
I(t) \simeq I_n(t) = \sum_{i=0}^n  a2_i~_c\psi_i(t),\\
R(t) \simeq R_n(t) = \sum_{i=0}^n  a3_i~_c\psi_i(t),\\
C(t) \simeq C_n(t) = \sum_{i=0}^n  a4_i~_c\psi_i(t),
\end{eqnarray}
for Influenza model. Similarly, we define it's derivative according to the Eq.(\ref{eqder}).
By placing the solution functions in the \ref{HIVMODEL} and \ref{SIRCMODEL} equations, we form the residual functions for each system of equations as follows.
\begin{eqnarray}
&Res_1(t) = \frac{d}{dt}T_n(t)+\alpha T_n(t)-rT_n(t)(1-\frac{T_n(t)+I_n(t)}{T_{max}})+kV_n(t)T_n(t)-s,\\
&Res_2(t) = \frac{d}{dt}I_n(t) - k V_n(t)T_n(t)+\beta I_n(t),\\
&Res_3(t) = \frac{d}{dt}V_n(t)-n\beta I_n(t)+\gamma V_n(t),
\end{eqnarray}
for HIV and
\begin{eqnarray}
&Res_1(t) = \frac{d}{dt}S_n(t) - \mu(1-S_n(t))+\beta S_n(t)I_n(t)+\gamma C_n(t),\\
&Res_2(t) = \frac{d}{dt}I_n(t) - \beta S_n(t)I_n(t) - \sigma\beta C_n(t)I_n(t) + (\mu + \theta)I_n(t),\\
&Res_3(t) = \frac{d}{dt}R_n(t) -(1-\sigma)\beta C_n(t)I_n(t)-\theta I_n(t)+(\mu+\delta)R_n(t),\\
&Res_4(t) = \frac{d}{dt}C_n(t) - \delta R_n(t)+\beta C_n(t)I_n(t)+(\mu+\gamma)C_n(t),
\end{eqnarray}
for Influenza.\\
We will simulate the problem in the domain (0,~1], and for this purpose we will choose the $n$ points $t=t_1,\cdots, t_n$  within the domain, where we have used equidistant points.
By placing the $n$ points in the residual functions and adding the initial conditions (IC\#) as follows
\begin{eqnarray}
IC1_{HIV} = T_n(0)-T_0,\\
IC2_{HIV} = I_n(0) - I_0,\\
IC3_{HIV} = V_n(0)-V_0,\\
IC1_{Influenza} = S_n(0)-S_0,\\
IC2_{Influenza} = I_n(0)-I_0,\\
IC3_{influenza} = R_n(0)-R_0,\\
IC4_{influenza} = C_n(0)-C_0.
\end{eqnarray}
We obtain $3(n+1)$ algebraic nonlinear equations for HIV model , as well as $4(n+1)$ algebraic nonlinear equations for Influenza model as follows

\begin{eqnarray}
&RES_{HIV} = (Res_1(t_1), \cdots, Res_1(t_n), Res_2(t_1), \cdots, Res_2(t_n), Res_3(t_1), \cdots, Res_3(t_n), IC1_{HIV}, IC2_{HIV}, IC3_{HIV} )_{3(n+1)},\\
&RES_{Influenza} = (Res_1(t_1), \cdots, Res_1(t_n), Res_2(t_1), \cdots, Res_2(t_n), Res_3(t_1), \cdots, Res_3(t_n),\nonumber\\ 
&Res_4(t_1), \cdots, Res_4(t_n), IC1_{Influenza}, IC2_{Influenza}, IC3_{Influenza}, IC4_{Influenza} )_{4(n+1)},
\end{eqnarray}
and finally for obtaining unknown coefficients, we solve these algebraic equations by Newton-Raphson method.
\subsection{Genetic algorithm to find optimal shape parameter}
GA as a meta-heuristic approach employed for optimization and finding the optimal parameter in problems. In fact, solving a problem by GA includes designing some functions and subroutines which be fired in each iteration (evolution). The main required functions and subroutines are Fitness function, Selection, Crossover, and mutation. However, more detailed explanation of GA is as follows:
\begin{enumerate}
\item \textbf{Generating an initial population (chromosomes)}: The algorithm utilizes a population-based structure to solve the problem. Thus it is necessary to pick out an initial population from the solution domain and start the evolution. Generating the initial population is usually done by a uniformly random distribution. The commands "\textbf{sample('Uniform'($\Omega_a$, $\Omega_b$),\#points)}" from the library "\textbf{Statistics}"  of Maple and "\textbf{rand(\#points)}" in Matlab generate the mentioned population.
\item \textbf{Fitness function}: In fitness function, the competency of each chromosome is investigated. The fitness of each individual is typically a numerical value. According to the nature of the problem, we assume that the minimum cost is zero and define our fitness function as:
$$\exp(\frac{1}{1+\Theta})$$
where $\Theta$ is ASN2R ($\frac{\sum_{i=1}^3 \|Res_i\|_2}{3}$) in HIV problem and ARE ($\frac{\sum_{i=1}^N |u_i-u_{RKF}|}{N}$) in SIRC model.

\item \textbf{Parental selection}: GA is an iterative process, with the population in each iteration called a generation. The more fit individuals (parents) are stochastically selected from the current population, and modified (recombined and possibly randomly mutated) to form a new generation (children). Then the new generation of candidate solutions are used in the next iteration of the algorithm. Our method for selecting parents is based on the fitness function and Roulette wheel technique (RWT). In RWT, the chance of an individual to be chosen as a parent has a direct relationship with its fitness value.
\item \textbf{Crossover}: When two individuals are selected as parents, the crossover subroutines combine them to produce a new individual (their child). In this regard, we define a crossover formula called "\textbf{Pseudo-combination}" (PCF). The PCF produces a child based on the value and fitness of its both parents. We define PCF as follows:
$$d = \frac{a+b}{2}+sign(b-a)(|b-\frac{a+b}{2}+sign(b-a)\varepsilon|)\frac{|f(a)|^\alpha-|f(b)|^\alpha}{|f(a)|^\alpha+|f(b)|^\alpha}$$
where $d, a, b, \varepsilon$ and $\alpha$ parameters are child, first parent, second parent, outer limit and strongly inclination, respectively.
The convergence of this PCF can be shown by taking the limit under different conditions as follows
\begin{enumerate}
\item If $|f(b)|^\alpha<|f(a)|^\alpha$ and $a<b$  so
\begin{eqnarray}
&\lim_{|f(b)|\rightarrow 0}d = \frac{a+b}{2}+sign(b-a)(|b-\frac{a+b}{2}+sign(b-a)\varepsilon |)\frac{|f(a)|^\alpha}{|f(a)|^\alpha}=\nonumber\\
&\frac{a+b}{2}+sign(b-a)(|b-\frac{a+b}{2}+sign(b-a)\varepsilon|)\nonumber
\end{eqnarray}
because $a<b$ so $b-\frac{a+b}{2}>0$ 
\begin{eqnarray}
&\frac{a+b}{2}+sign(b-a)(|b-\frac{a+b}{2}+sign(b-a)\varepsilon|)=\nonumber\\
&\frac{a+b}{2}+(b-\frac{a+b}{2}+\varepsilon)=b+\varepsilon\nonumber\\
&\lim_{|f(b)|\rightarrow 0}d=b+\varepsilon\nonumber.
\end{eqnarray}
\item If $|f(b)|^\alpha<|f(a)|^\alpha$ and $a>b$  so $b-\frac{a+b}{2}<0$ 
\begin{eqnarray}
&\frac{a+b}{2}+sign(b-a)(|b-\frac{a+b}{2}+sign(b-a)\varepsilon|)=\nonumber\\
&\frac{a+b}{2}-(-b+\frac{a+b}{2}+\varepsilon)=b-\varepsilon\nonumber\\
&\lim_{|f(b)|\rightarrow 0}d=b-\varepsilon\nonumber.
\end{eqnarray}

\item If $|f(b)|^\alpha>|f(a)|^\alpha$ and $a<b$ so
\begin{eqnarray}
&\lim_{|f(a)|\rightarrow 0}d = \frac{a+b}{2}+sign(b-a)(|b-\frac{a+b}{2}+sign(b-a)\varepsilon|)\frac{-|f(b)|^\alpha}{|f(b)|^\alpha}=\nonumber\\
&\frac{a+b}{2}-sign(b-a)(|b-\frac{a+b}{2}+sign(b-a)\varepsilon|)\nonumber
\end{eqnarray}
because $a<b$ so $b-\frac{a+b}{2}>0$ 
\begin{eqnarray}
&\frac{a+b}{2}-sign(b-a)(|b-\frac{a+b}{2}+sign(b-a)\varepsilon|)=\nonumber\\
&\frac{a+b}{2}-(b-\frac{a+b}{2}+\varepsilon)=a-\varepsilon\nonumber\\
&\lim_{|f(a)|\rightarrow 0}d=a-\varepsilon\nonumber.
\end{eqnarray}
\item If $|f(b)|^\alpha>|f(a)|^\alpha$ and $a>b$  so $b-\frac{a+b}{2}<0$ 
\begin{eqnarray}
&\frac{a+b}{2}-sign(b-a)(|b-\frac{a+b}{2}+sign(b-a)\varepsilon|)=\nonumber\\
&\frac{a+b}{2}+(-b+\frac{a+b}{2}+\varepsilon)=a+\varepsilon\nonumber\\
&\lim_{|f(a)|\rightarrow 0}d=a+\varepsilon\nonumber.
\end{eqnarray}
\item If $|f(b)|^\alpha=|f(a)|^\alpha$ so 
\begin{eqnarray}
\lim_{|f(a|b)|\rightarrow 0}d = \frac{a+b}{2}.\nonumber
\end{eqnarray}
\end{enumerate}
\item \textbf{mutation}: In mutation operation, some chromosomes are chosen randomly (according to the mutation rate) and one digit of each chromosome is replaced with a random digit. Considering elitism, we guard top three chromosomes (based on their fitness) against mutation.
\end{enumerate}
Algorithm (\ref{GAKANSA}) presents a general form of the proposed GA.
\begin{algorithm}
\caption{GA-RBF method }
\label{GAKANSA}
 Initial random feasible population: $pop \leftarrow \{c_1, c_2, \cdots, c_N\}$\\
\textbf{WHILE} (iteration condition)\\
\quad RBF method computes ASN2R (or ARE) for population\\
\quad Fitness $\leftarrow \exp(\frac{1}{1+ASN2R(or ARE)})$\\
\quad Sorting the population : $\forall c_i,c_j \in pop, i<j \iff Fitness(c_i)>Fitness(c_j)$\\
\quad Elitism : $pop' \leftarrow \{c_1, c_2, \cdots, c_{elit}\}$\\
\quad \textbf{FOR} $i~=~elit+1 ~~ to ~~ N$ \\
\qquad $ \{p_1, p_2\} \leftarrow ParentalSelection(pop)$\\
\qquad $c'_i \leftarrow PCF(p_1,p_2)$\\
\qquad $c''_i \leftarrow Mutation(c'_i)$\\
\qquad $pop' \leftarrow pop' \cup \{c''_i\}$\\
\quad \textbf{END FOR}\\
\quad $pop \leftarrow pop'$\\
\textbf{END WHILE}
 
\end{algorithm}
\section{Result}\label{B}

In this section, we set $\varepsilon$=0.02, $\alpha$ = 0.016 $mutation$=0.2 and $elit$=3, and solve the HIV(\ref{HIVMODEL}) and Influenza SIRC(\ref{SIRCMODEL}) models. We used Maple 2015 for solving these models. The hardware configuration was as follows: \\
OS : Windows 10 (64bit)\\
CPU : Corei7 2.8 GHZ\\
RAM : 12 GB DDR3.

\subsection{HIV}
\begin{figure}[htbp!]
    \centering
    \subfigure[T(t)]
    {
        \includegraphics[width=0.3\textwidth]{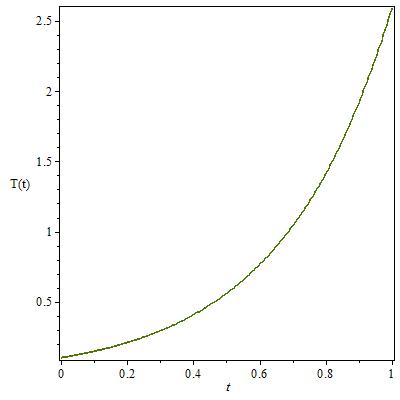}
        \label{PlT}
    }
    \\
    \subfigure[I(t) ]
    {
        \includegraphics[width=0.3\textwidth]{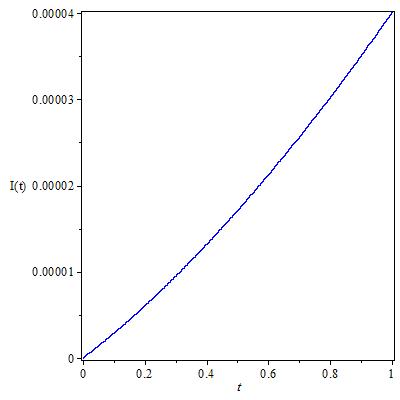}
        \label{PlI}
    }
    ~~~
    \subfigure[V(t)]
    {
        \includegraphics[width=0.3\textwidth]{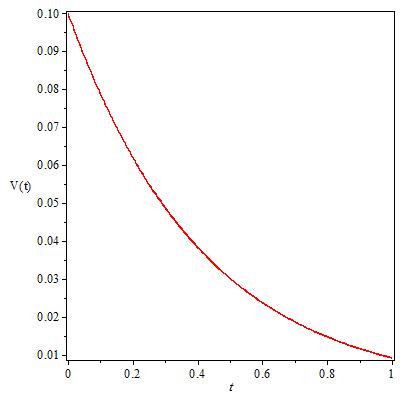}
        \label{PlV}
    }
    \caption{Plots of $T(t), I(t), V(t)$ for $N=20$}
    \label{fig:PlotsHIV}
\end{figure}

\begin{figure}[htbp!]
    \centering
    \subfigure[$Res_{T(t)}$]
    {
        \includegraphics[width=0.3\textwidth]{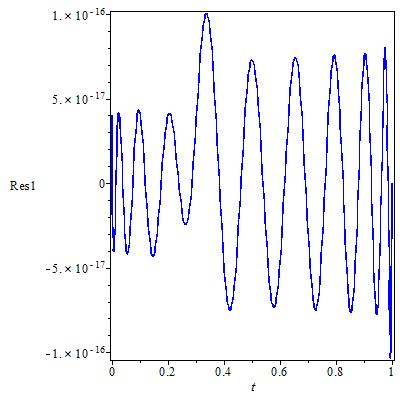}
        \label{PlT}
    }
    \\
    \subfigure[$Res_{I(t)}$]
    {
        \includegraphics[width=0.3\textwidth]{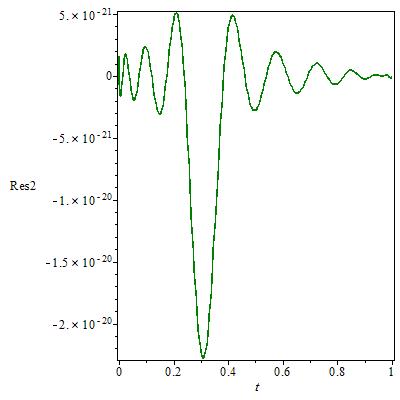}
        \label{PlI}
    }
    ~~~
    \subfigure[$Res_{V(t)}$]
    {
        \includegraphics[width=0.3\textwidth]{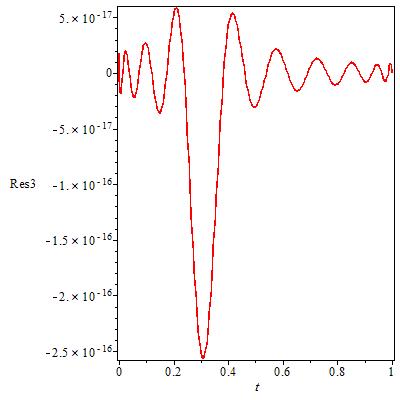}
        \label{PlV}
    }
    \caption{Plots of  residual $T(t), I(t), V(t)$ for $N=20$}
    \label{fig:Plots}
\end{figure}

In HIV model, we approximate target functions with the classic Gaussian function and apply the average of residual functions to the fitness function. Figures \ref{fig:PlotsHIV} and \ref{fig:Plots} show given target function and residual function plots from 20 collocation points for T(t), I(t) and V(t). 

Tables \ref{Tres},\ref{Ires} and \ref{Vres} illustrates the comparison of the presented method (with 20 collocation points) with approximated results of Bessel Collocation method (BCM)\cite{yuzbacsi2012numerical}, Runge-Kutta method (RKM)\cite{yuzbacsi2012numerical}, Homotopy Decomposition method (HDM)\cite{atangana2014computational} and Wavelet Legendre method (WLM)\cite{venkatesh2016new}. The results show that the presented method and Runge-Kutta method are equal up to eight decimal places.
\begin{table}[h!]
\caption{Numerical results for $T(t)$}
\label{Tres}
\begin{tabular}{|c|c|c|c|c|c|}
\hline 
t & BCM & RKM & HDM & LWM & Present method for N=20 \\ 
\hline
0.2 & 0.2038616561 & 0.2088080833 & 0.2088072731 & 0.2088073215 & 0.2088080843 \\ 
0.4 & 0.3803309335 & 0.4062405393 & 0.4061052625 & 0.4061245634 & 0.4062405427 \\ 
0.6 & 0.6954623767 & 0.7644238890 & 0.7611467713 & 0.7641476415 & 0.7644238985 \\ 
0.8 & 1.2759624442 & 1.4140468310 & 1.3773198590 & 1.3977746217 & 1.4140468518 \\ 
1.0 & 2.3832277428 & 2.5915948020 & 2.3291697610 & 2.5571462314 & 2.5915948516 \\ 
\hline 
\end{tabular} 
\end{table}

\begin{table}[h!]
\caption{Numerical results for $I(t)$}
\label{Ires}
\begin{tabular}{|c|c|c|c|c|c|}
\hline 
t & BCM & Runge-Kutta & HDM & LWM & Present method for N=20 \\ 
\hline 
0.2 & 0.6247872e-5 & 0.6032702e-5 & 6.0327072e-5 & 0.6032704e-5 & 0.6032702e-5 \\ 
0.4 & 0.1293552e-4 & 0.1315834e-4 & 1.3159161e-4 & 0.1316784e-4 & 0.1315834e-4 \\ 
0.6 & 0.2035267e-4 & 0.2122378e-4 & 2.1268368e-4 & 0.2112628e-4 & 0.2122378e-4 \\ 
0.8 & 0.2837302e-4 & 0.3017741e-4 & 3.0069186e-4 & 0.2998139e-4 & 0.3017742e-4 \\  
1.0 & 0.3690842e-4 & 0.4003781e-4 & 3.9873654e-4 & 0.3287654e-4 & 0.4003781e-4 \\ 
\hline 
\end{tabular} 
\end{table}

\begin{table}[h!]
\caption{Numerical results for $V(t)$}
\label{Vres}
\begin{tabular}{|c|c|c|c|c|c|}
\hline 
t & BCM & Runge-Kutta & HDM & LWM & Present method for N=20 \\
\hline 
0.2 & 0.0618799185 & 0.0618798433 & 0.0618799602 & 0.0618799076 & 0.0618798432 \\ 
0.4 & 0.0382949349 & 0.0382948878 & 0.0383132488 & 0.0383234157 & 0.0382948877 \\ 
0.6 & 0.0237043186 & 0.0237045501 & 0.0243917434 & 0.0238109873 & 0.0237045500 \\ 
0.8 & 0.0146795698 & 0.0146803637 & 0.0099672189 & 0.0162138976 & 0.0146803636 \\ 
1.0 & 0.0237043186 & 0.0091008450 & 0.0033050764 & 0.0160504423 & 0.0091008449 \\ 
\hline 
\end{tabular} 
\end{table}

\begin{figure}[htbp!]
    \centering
    \subfigure[ N=5]
    {
        \includegraphics[width=0.3\textwidth]{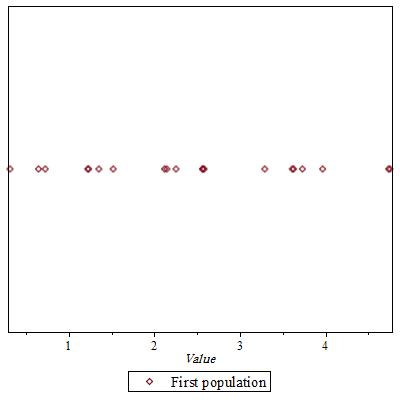}
        \label{IP5}
    }
    ~~~
      \subfigure[ N=10]
    {
        \includegraphics[width=0.3\textwidth]{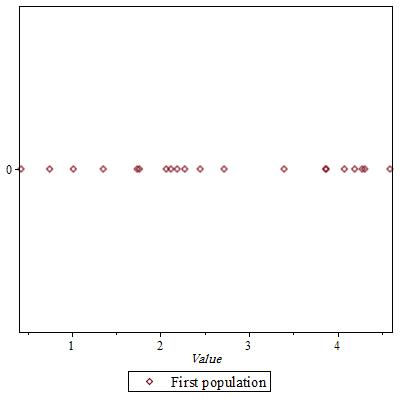}
        \label{IP10}
    }
    \\
    \subfigure[ N=15]
    {
        \includegraphics[width=0.3\textwidth]{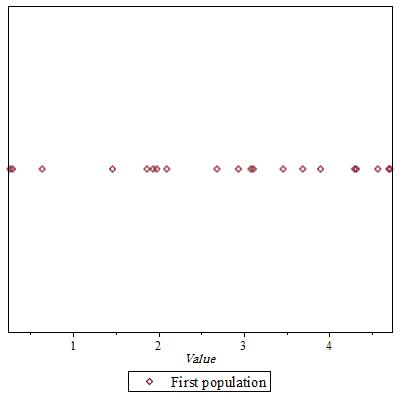}
        \label{IP15}
    }
    ~~~
    \subfigure[N=20]
    {
        \includegraphics[width=0.3\textwidth]{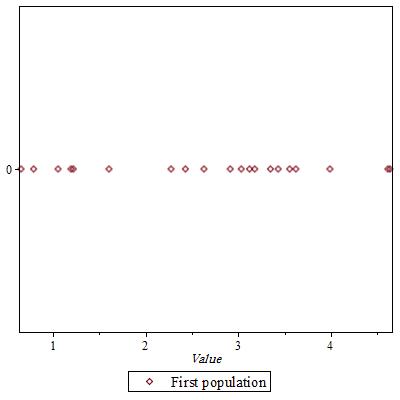}
        \label{IP20}
    }
    \caption{Initial population for SP domain}
    \label{fig:IPSP}
\end{figure}

\begin{figure}[htbp!]
    \centering
    \subfigure[ N=5]
    {
        \includegraphics[width=0.3\textwidth]{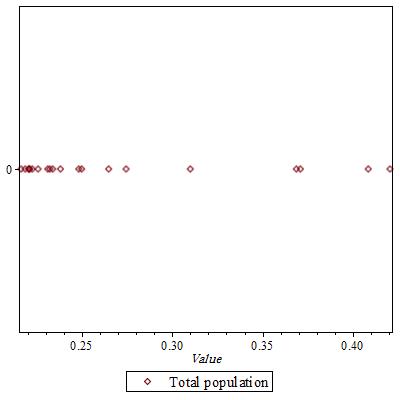}
        \label{TP5}
    }
    ~~~
      \subfigure[ N=10]
    {
        \includegraphics[width=0.3\textwidth]{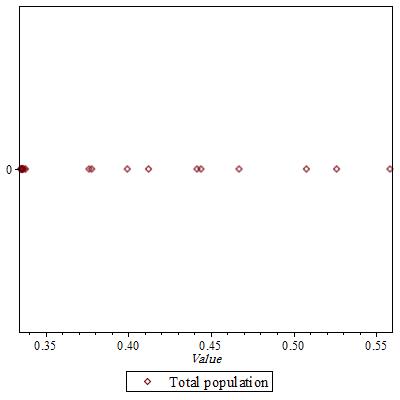}
        \label{TP10}
    }
    \\
    \subfigure[ N=15]
    {
        \includegraphics[width=0.3\textwidth]{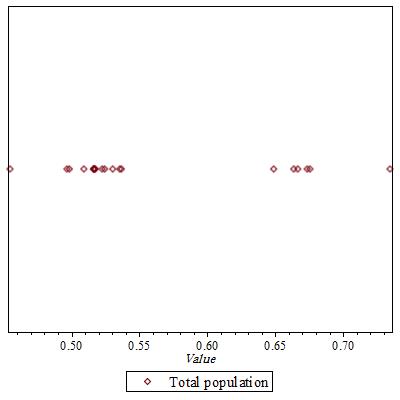}
        \label{TP15}
    }
    ~~~
    \subfigure[ N=20]
    {
        \includegraphics[width=0.3\textwidth]{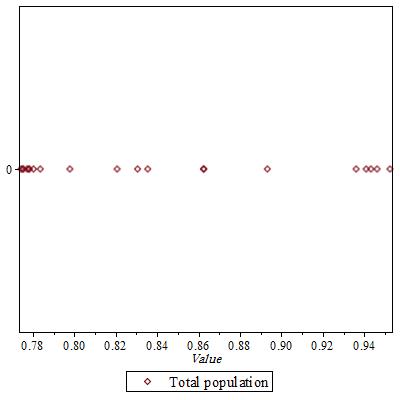}
        \label{TP20}
    }
    \caption{Total population for SP domain in latest iteration}
    \label{fig:TPSP}
\end{figure}

Figures \ref{fig:IPSP} and \ref{fig:TPSP} show the residual functions. We used the uniform distribution library of Maple and generate 20 chromosomes as GA initial population in search domain (0.1 , 5). The GA population collected on the smaller range after 20 iterations:\\
5 collocation points (0.1 , 5)$\longrightarrow$(0.15 , 0.45)\\
10 collocation points (0.1 , 5)$\longrightarrow$ (0.3 , 0.59)\\
15 collocation points (0.1 , 5)$\longrightarrow $(0.45 , 0.75)\\
20 collocation points (0.1 , 5)$\longrightarrow $(0.74 , 0.95).\\ 

\begin{figure}[htbp!]
    \centering
    \subfigure[$N=5$]
    {
        \includegraphics[width=0.3\textwidth]{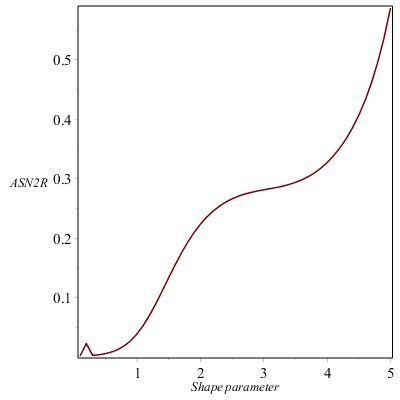}
        \label{PlI}
    }
    ~~~
    \subfigure[$N=15$]
    {
        \includegraphics[width=0.3\textwidth]{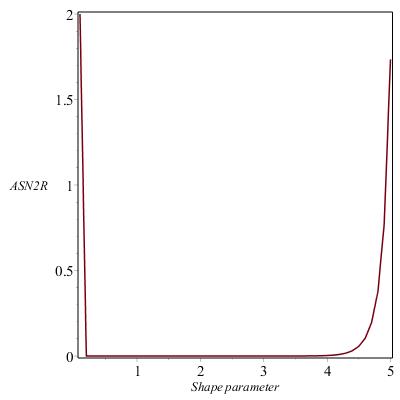}
        \label{PlV}
    }
    \caption{ASN2R condition based on Shape parameter}
    \label{SHAPE}
\end{figure}

Figure \ref{SHAPE} shows the condition of ASN2R based on the SP in the domain (0.1, 5). The optimal SP for 5 collocation points is in the domain (0.12, 0.35) and for 15 collocation points is in the domain (0.2, 0.8).

Obviously, by increasing the number of iterations in GA, in the case of the uniqueness of the optimal point, the final range will be limited again.
Table \ref{Comp}  represents changes in the results and residuals by changing the number of collocation points. It can be seen that the results remained stable in 15 and 20 points. 

\begin{table}[htbp!]
\caption{Numerical comparison  for $N=5, 10, 15, 20$ in best SP}
\label{Comp}
\centering \begin{tabular}{|ccccccccc|}
\hline 
t & N=5 & Res & N=10 & Res & N=15 & Res & N=20 & Res \\ 
\hline 
 & c=0.21635819 &  & c=0.33489998 & T(t)& c=0.51637831&  & c=0.77428513 &  \\ 
\hline 
0.2 & 0.2141496490 & 3.86e-03 & 0.2088094672 & 7.82e-06 & 0.2088080843 & 2.60e-13 & 0.2088080843 & 3.96e-17 \\ 
0.4 & 0.3996040883 & 1.31e-01 & 0.4062432618 & 7.24e-07 & 0.4062405427 & 1.68e-13 & 0.4062405427 & 4.87e-17 \\ 
0.6 & 0.7116524148 & 1.34e-01 & 0.7644289795 & 4.05e-07 & 0.7644238985 & 1.55e-13 & 0.7644238985 & 4.65e-17 \\ 
0.8 & 1.2973641582 & 4.31e-03 & 1.4140559835 & 6.47e-07 & 1.4140468518 & 3.61e-13 & 1.4140468518 & 7.23e-17 \\ 
1.0 & 2.3892259227 & 5.66e-92 & 2.5916114131 & 1.94e-84 & 2.5915948516 & 3.17e-80 & 2.5915948516 & 8.88e-78 \\ 
\hline 
 &  &  &  & I(t)&  &  &  &  \\ 
\hline 
0.2 & 0.6114687e-5 & 2.27e-09 & 0.6032719e-5 & 2.46e-11 & 0.6032702e-5 & 3.42e-18 & 0.6032702e-5 & 4.46e-21 \\ 
0.4 & 0.1337344e-4 & 8.34e-07 & 0.1315841e-4 & 5.26e-14 & 0.1315834e-4 & 6.28e-18 & 0.1315834e-4 & 3.71e-21 \\ 
0.6 & 0.2134401e-4 & 1.13e-06 & 0.2122391e-4 & 6.30e-13 & 0.2122378e-4 & 5.06e-18 & 0.2122378e-4 & 1.09e-21 \\ 
0.8 & 0.2987347e-4 & 3.22e-08 & 0.3017760e-4 & 1.60e-12 & 0.3017742e-4 & 7.01e-19 & 0.3017742e-4 & 6.41e-22 \\  
1.0 & 0.3908961e-4 & 6.10e-96 & 0.4003806-4 & 4.41e-89 & 0.4003781e-4 & 9.99e-85 & 0.4003781e-4 & 1.45e-81 \\
\hline 
 &  &  &  & V(t)&  &  &  &  \\  
 \hline
0.2 & 0.0618187051 & 6.80e-05 & 0.0618798524 & 6.92e-08 & 0.0618798432 & 4.57e-14 & 0.0618798432 & 5.14e-17 \\ 
0.4 & 0.0384572762 & 2.28e-03 & 0.0382948935 & 6.31e-09 & 0.0382948877 & 7.37e-14 & 0.0382948877 & 4.08e-17 \\ 
0.6 & 0.0242145096 & 2.31e-03 & 0.0237045544 & 3.49e-09 & 0.0237045500 & 7.32e-14 & 0.0237045500 & 1.22e-17 \\ 
0.8 & 0.0151695545 & 7.31e-05 & 0.0146803662 & 5.55e-09 & 0.0146803636 & 4.63e-14 & 0.0146803636 & 1.00e-17 \\ 
1.0 & 0.0093128300 & 3.10e-93 & 0.0091008467 & 2.11e-86 & 0.0091008449 & 6.50e-82 & 0.0091008449 & 4.10e-79 \\ 
\hline 
\end{tabular} 
\end{table}

\begin{figure}[htbp!]
    \centering
    \subfigure[ N=5 ]
    {
        \includegraphics[width=0.3\textwidth]{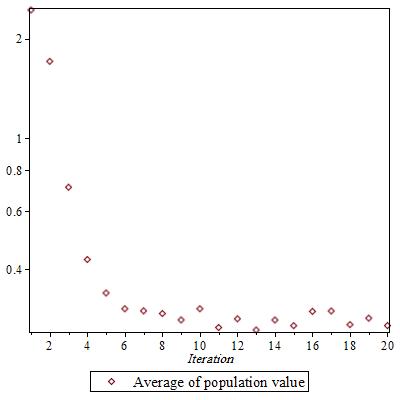}
        \label{APV5}
    }
    ~~~
      \subfigure[ N=10]
    {
        \includegraphics[width=0.3\textwidth]{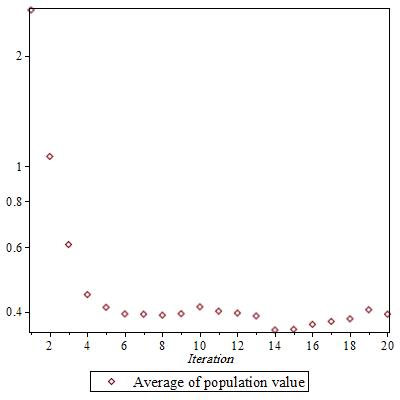}
        \label{APV10}
    }
    \\
    \subfigure[ N=15]
    {
        \includegraphics[width=0.3\textwidth]{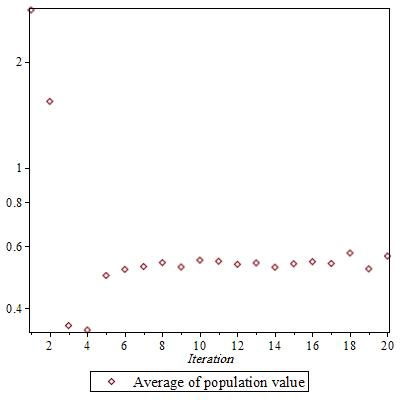}
        \label{APV15}
    }
    ~~~
    \subfigure[ N=20]
    {
        \includegraphics[width=0.3\textwidth]{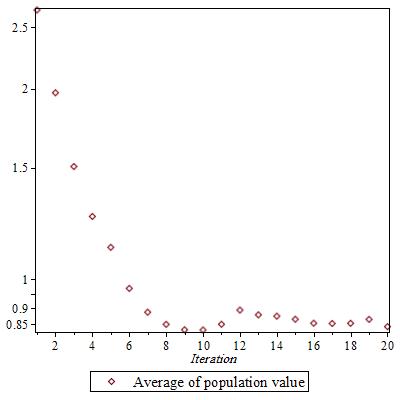}
        \label{APV20}
    }
    \caption{Average of population value in iterations}
    \label{fig:APVSP}
\end{figure}

\begin{figure}[htbp!]
    \centering
    \subfigure[ N=5 ]
    {
        \includegraphics[width=0.3\textwidth]{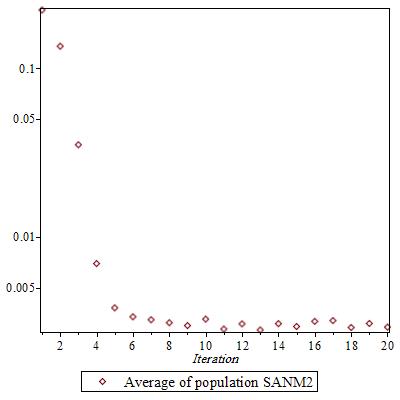}
        \label{APF5}
    }
    ~~~
      \subfigure[ N=10]
    {
        \includegraphics[width=0.3\textwidth]{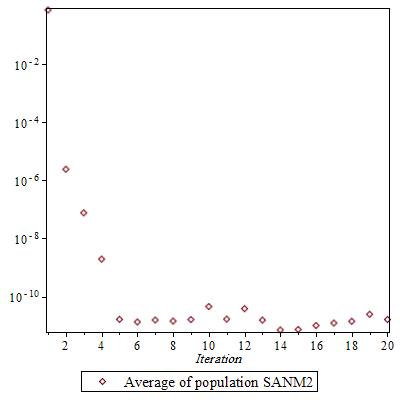}
        \label{APF10}
    }
    \\
    \subfigure[ N=15]
    {
        \includegraphics[width=0.3\textwidth]{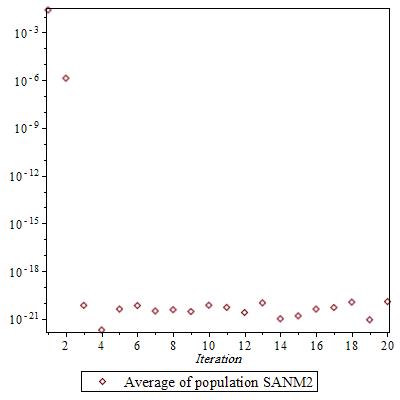}
        \label{APF15}
    }
    ~~~
    \subfigure[ N=20]
    {
        \includegraphics[width=0.3\textwidth]{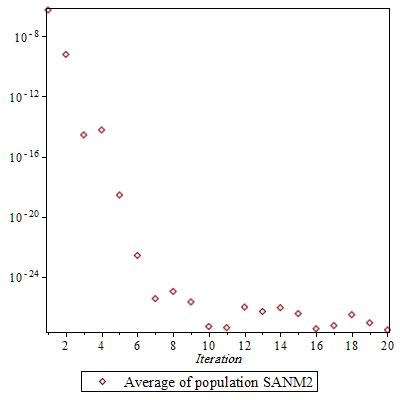}
        \label{APF20}
    }
    \caption{Average of population fitness in iterations}
    \label{fig:APFSP}
\end{figure}

 In Fig. (\ref{fig:APVSP}) and Fig. (\ref{fig:APFSP}) display the average of value population (AVP) and the average of the residual population (ARP). After a number of steps, the AVP tended to a nonzero value, likewise, the ARP disposed to zero which indicates the convergence and productivity of our Genetic strategy.

\subsection{Influenza}
\begin{figure}
\centering \includegraphics[width=0.4\textwidth]{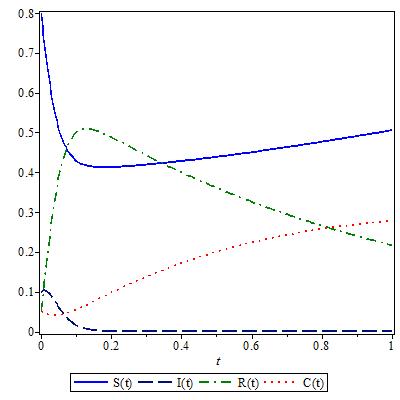}
\caption{Plots for SIRC model}
\label{SIRCPLOT}
\end{figure}

\begin{figure}[htbp!]
    \centering
    \subfigure[S(t) ]
    {
        \includegraphics[width=0.3\textwidth]{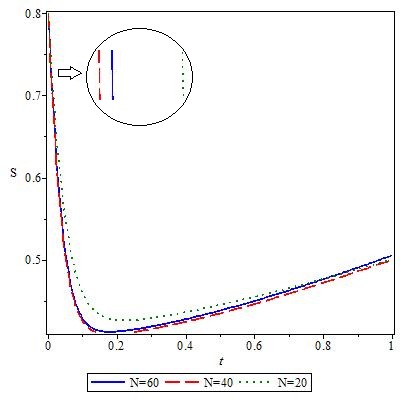}
        \label{ST}
    }
    ~~~
      \subfigure[I(t)]
    {
        \includegraphics[width=0.3\textwidth]{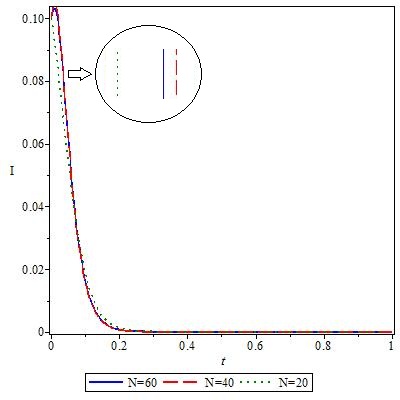}
        \label{I(t)}
    }
    \\
    \subfigure[R(t)]
    {
        \includegraphics[width=0.3\textwidth]{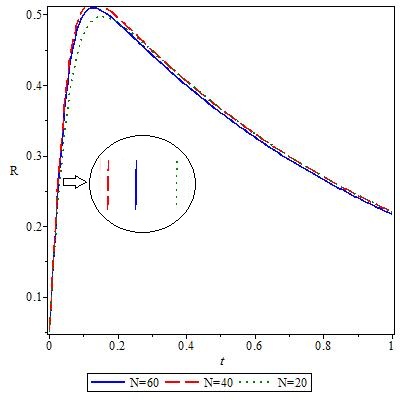}
        \label{RT}
    }
    ~~~
    \subfigure[C(t)]
    {
        \includegraphics[width=0.3\textwidth]{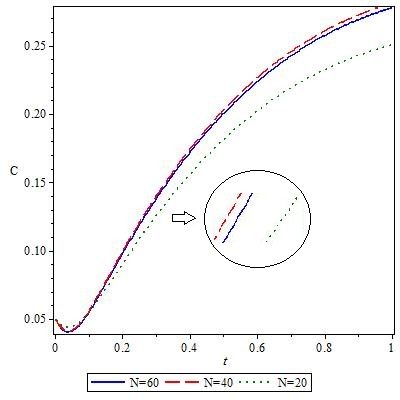}
        \label{CT}
    }
    \caption{Gained plots for $N=20, 40, 60$}
    \label{fig:SIRC}
\end{figure}

\begin{figure}[htbp!]
    \centering
    \subfigure[$N=60$]
    {
        \includegraphics[width=0.3\textwidth]{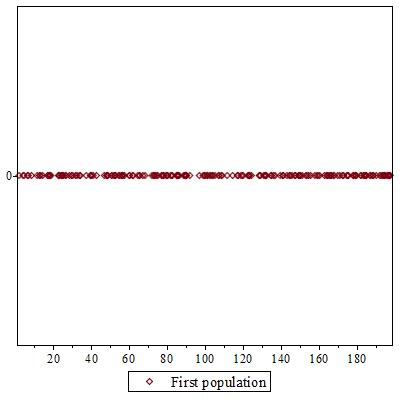}
        \label{AIPlT}
    }
    \\
    \subfigure[$N=40$]
    {
        \includegraphics[width=0.3\textwidth]{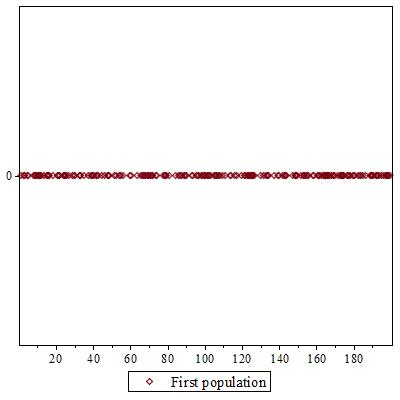}
        \label{AIPlI}
    }
    ~~~
    \subfigure[$N=20$]
    {
        \includegraphics[width=0.3\textwidth]{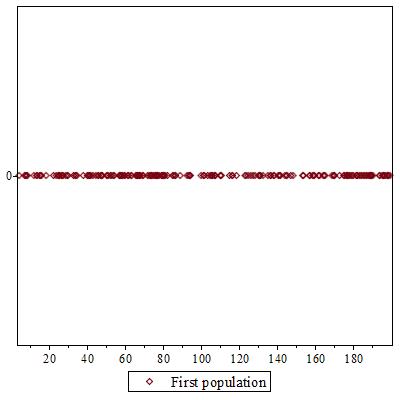}
        \label{AIPlV}
    }
    \caption{Initial population for SP domain}
    \label{fig:SIRCATP}
\end{figure}

\begin{figure}[htbp!]
    \centering
    \subfigure[$N=60$]
    {
        \includegraphics[width=0.3\textwidth]{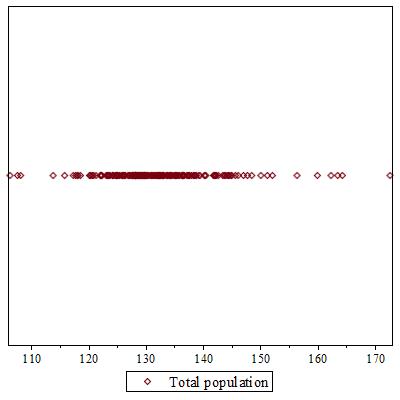}
        \label{ATPlT}
    }
    \\
    \subfigure[$N=40$]
    {
        \includegraphics[width=0.3\textwidth]{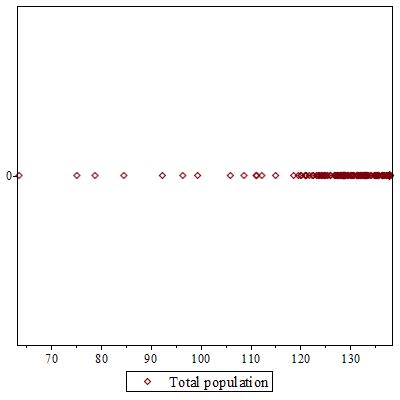}
        \label{ATPlI}
    }
    ~~~
    \subfigure[$N=20$]
    {
        \includegraphics[width=0.3\textwidth]{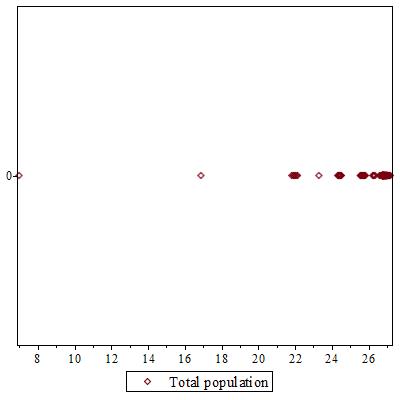}
        \label{ATPlV}
    }
    \caption{Total population for SP domain}
    \label{fig:SIRCTTP}
\end{figure}

In this model, we applied the Gaussian function with a minor change as follows
\begin{eqnarray}
\label{Gwide}
 _c\psi = \exp(-\eta r^2)~~~~~~~~\eta=\sqrt(c^2),
\end{eqnarray}
Moreover, the GA population is 200 randomly selected points from the real domain (1, 200). We used Maple's DSOLVE tool for calculating the fitness of chromosomes and compared our results with Runge-Kutta-Fehlberg (RKF) method. Four target functions have been obtained from 60 collocation points whose plots are shown in Fig.(\ref{SIRCPLOT}). In addition, in Fig. (\ref{fig:SIRC}), treatment of functions for 20, 40 and 60 collocation points are considered. Initial and total population are presented in Fig. (\ref{fig:SIRCATP}) and Fig. (\ref{fig:SIRCTTP}).

\begin{figure}[htbp!]
    \centering
    \subfigure[$N=20$]
    {
        \includegraphics[width=0.3\textwidth]{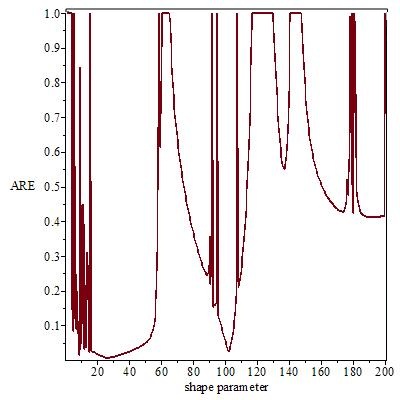}
        \label{PlI}
    }
    ~~~
    \subfigure[$N=40$]
    {
        \includegraphics[width=0.3\textwidth]{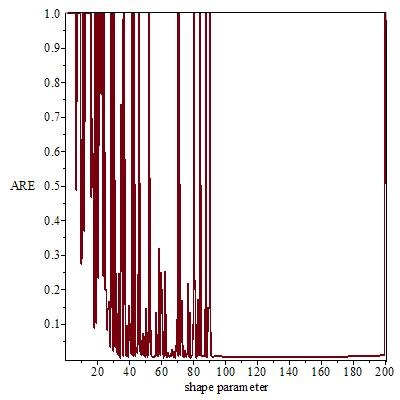}
        \label{PlV}
    }
    \caption{ARE condition based on Shape parameter}
    \label{SHAPEST}
\end{figure}

\begin{figure}[htbp!]
    \centering
    \subfigure[$N=60$]
    {
        \includegraphics[width=0.3\textwidth]{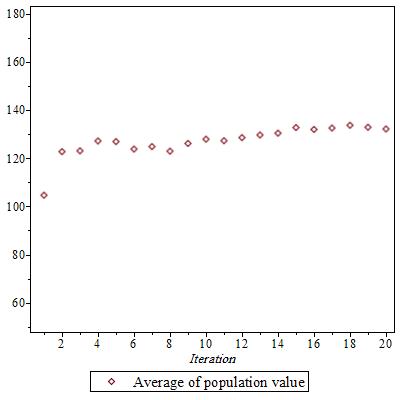}
        \label{APlT}
    }
    \\
    \subfigure[$N=40$]
    {
        \includegraphics[width=0.3\textwidth]{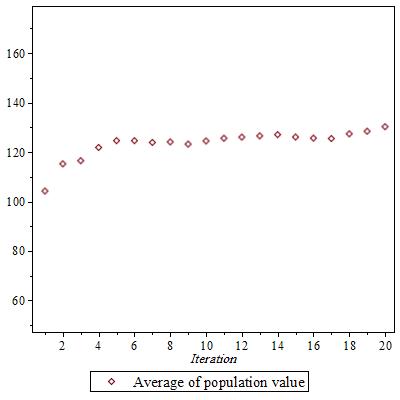}
        \label{APlI}
    }
    ~~~
    \subfigure[$N=20$]
    {
        \includegraphics[width=0.3\textwidth]{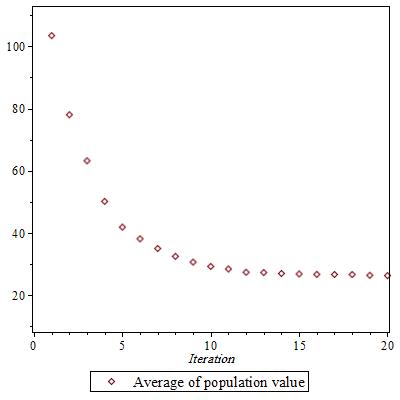}
        \label{APlV}
    }
    \caption{Average of population value in  iterations}
    \label{fig:SIRCAIP}
\end{figure}

 When we set the number of collocation points to 20, 40 and 60, the final generation collected in ranges (22, 28), (100, 140) and (120, 150)  respectively.  Furthermore, convergence of APN to a nonzero value is searchable in Fig.(\ref{fig:SIRCAIP}). Figure \ref{SHAPEST}  shows the condition of ARE based on the SP selected from the domain (1, 200). For 20 collocation points, the optimal SP is in the domain (20, 30) and for 40 collocation points is in the domain (100, 160).
 
Eventually, Tab.(\ref{InfTab}) displays the comparison of the proposed method (for 20, 40 and 60 collocation points) with RKF method. It shows that our results converge to the RKF method by increasing the number of collocation points.

\begin{table}
\caption{Runge-Kutta-Fehlberg and RBF method results with $N=20, 40, 60$}
\begin{tabular}{|c|c|c|c|c|c|}
\hline 
       & RKFM & Presented method 20N & Presented method 40N & Presented method 60N & Relative error 60N \\ 
\hline 
  &    &     & S(t) &    &   \\ 
\hline 
c & - & 26.800747761 & 137.9078869 & 123.44438040 & - \\ 
\hline
0.2 & 0.4129967652 & 0.4270195495 & 0.4098921081 & 0.4128015008 & 1.952644e-04 \\ 
0.4 & 0.4280199934 & 0.4363880004 & 0.4246103505 & 0.4278128844 & 2.071090e-04 \\ 
0.6 & 0.4502002535 & 0.4549500702 & 0.4463055244 & 0.4499962534 & 2.040001e-04 \\ 
0.8 & 0.4765242244 & 0.4766921230 & 0.4719383037 & 0.4763283175 & 1.959069e-04 \\ 
1.0 & 0.5054118548 & 0.5003541451 & 0.5001004305 & 0.5052268175 & 1.850373e-04 \\ 
\hline 
\hline 
  &    &     & I(t) &    &   \\ 
\hline 
c & - & 26.800747761 & 137.9078869 & 123.44438040 & - \\ 
\hline
0.2 & 7.340828e-04 & 1.396989e-03 & 7.051238e-04 & 7.534934e-04 & 0.194106e-04 \\ 
0.4 & 1.675623e-06 & 8.708980e-06 & 1.566126e-06 & 1.806640e-06 & 1.310170e-07 \\ 
0.6 & 2.525609e-09 & 8.678259e-08 & 4.611322e-09 & -1.672634e-09 & 4.198243e-09 \\ 
0.8 & 9.855557e-10 & -3.67796e-07 & -2.23659e-09 & -1.066306e-09 & 2.0518617e-09 \\ 
1.0 & -7.72143e-10 & -2.21603e-05 & -1.428391e-09 & -3.324845e-08 & 3.2476307e-08 \\ 
\hline 
\hline 
  &    &     & R(t) &    &   \\ 
\hline 
c & - & 26.800747761 & 137.9078869 & 123.44438040 & - \\ 
\hline
0.2 & 0.4886071026 & 0.4886065768 & 0.4957432281 & 0.4884197981 & 1.873045e-04 \\ 
0.4 & 0.4001052933 & 0.4061157789 & 0.4061422515 & 0.3999667848 & 1.385085-e04 \\ 
0.6 & 0.3262752258 & 0.3313413846 & 0.3312018206 & 0.3261621289 & 1.130969e-04 \\ 
0.8 & 0.2660651578 & 0.2702079076 & 0.2700828333 & 0.2659730573 & 9.210050e-05 \\ 
1.0 & 0.2169661278 & 0.2203903869 & 0.2202819063 & 0.2168965849 & 6.954290e-05 \\  
\hline 
\hline 
  &    &     & C(t) &    &   \\ 
\hline 
c & - & 26.800747761 & 137.9078869 & 123.44438040 & - \\ 
\hline 
0.2 & 0.0976620493 & 0.9008003610 & 0.0990852949 & 0.0975773248 & 8.472450e-05 \\ 
0.4 & 0.1718730376 & 0.1556924837 & 0.1741092475 & 0.1717627291 & 1.103085e-04 \\ 
0.6 & 0.2235245180 & 0.2019667950 & 0.2263417601 & 0.2233978645 & 1.266535e-04 \\ 
0.8 & 0.2574106166 & 0.2323192884 & 0.2606042265 & 0.2572747238 & 1.358928e-04 \\ 
1.0 & 0.2776220180 & 0.2504052565 & 0.2810922415 & 0.2774868004 & 1.352176e-04 \\ 
\hline 
\end{tabular} 
\label{InfTab}
\end{table}
\section{Conclusion}
In this study we have proposed an approximation technique to solve biological equations.The method is based on the collocation method and Gaussian radial basis function. We used a Genetic strategy to overcome the challenge of searching optimal Shape parameters in RBF method. Additionally we tested ASN2R for HIV and ARE for SIRC model in fitness function and a new crossover formula called Pseudo-combination defined for using the considered GA. Finally, we showed that our approach is applicable and accurate for the solving system of the differential equations such as HIV and Influenza models.

\section*{Acknowledgments}
\newpage

The authors are very grateful to anonymous referees for carefully reading this paper and their comments and suggestions, which have improved the quality of the paper 

\bibliographystyle{abbrv}

\bibliography{BibFile}

\begin{thebibliography}{10}

\bibitem{abbasbandy2015shooting}
S.~Abbasbandy, B.~Azarnavid, and M.~S. Alhuthali.
\newblock A shooting reproducing kernel hilbert space method for multiple
  solutions of nonlinear boundary value problems.
\newblock {\em Journal of Computational and Applied Mathematics}, 279:293--305,
  2015.

\bibitem{afiatdoust2015optimal}
F.~Afiatdoust and M.~Esmaeilbeigi.
\newblock Optimal variable shape parameters using genetic algorithm for radial
  basis function approximation.
\newblock {\em Ain Shams Engineering Journal}, 6(2):639--647, 2015.

\bibitem{agusto2017mathematical}
F.~Agusto.
\newblock Mathematical model of ebola transmission dynamics with relapse and
  reinfection.
\newblock {\em Mathematical biosciences}, 283:48--59, 2017.

\bibitem{anderson1992infectious}
R.~M. Anderson and R.~M. May.
\newblock {\em Infectious diseases of humans: dynamics and control}.
\newblock Oxford university press, 1992.

\bibitem{atangana2014computational}
A.~Atangana and E.~F. Doungmo~Goufo.
\newblock Computational analysis of the model describing hiv infection of cd4+
  t cells.
\newblock {\em BioMed research international}, 2014, 2014.

\bibitem{atluri1998new}
S.~N. Atluri and T.~Zhu.
\newblock A new meshless local petrov-galerkin (mlpg) approach in computational
  mechanics.
\newblock {\em Computational mechanics}, 22(2):117--127, 1998.

\bibitem{azarnavid2015picard}
B.~Azarnavid, F.~Parvaneh, and S.~Abbasbandy.
\newblock Picard-reproducing kernel hilbert space method for solving
  generalized singular nonlinear lane-emden type equations.
\newblock {\em Mathematical Modelling and Analysis}, 20(6):754--767, 2015.

\bibitem{belytschko1995element}
T.~Belytschko, Y.~Lu, L.~Gu, and M.~Tabbara.
\newblock Element-free galerkin methods for static and dynamic fracture.
\newblock {\em International Journal of Solids and Structures},
  32(17-18):2547--2570, 1995.

\bibitem{bennett2014principles}
J.~Bennett, R.~Dolin, and M.~Blaser.
\newblock Principles and practice of infectious diseases, vol 1 elsevier health
  sciences.
\newblock {\em Philadelphia, PA.[Google Scholar]}, 2014.

\bibitem{berge2017simple}
T.~Berge, J.-S. Lubuma, G.~Moremedi, N.~Morris, and R.~Kondera-Shava.
\newblock A simple mathematical model for ebola in africa.
\newblock {\em Journal of biological dynamics}, 11(1):42--74, 2017.

\bibitem{chen2015adomian}
F.~Chen and Q.-Q. Liu.
\newblock Adomian decomposition method combined with pad{\'e} approximation and
  laplace transform for solving a model of hiv infection of cd4+ t cells.
\newblock {\em Discrete Dynamics in Nature and Society}, 2015, 2015.

\bibitem{colagrossi2003numerical}
A.~Colagrossi and M.~Landrini.
\newblock Numerical simulation of interfacial flows by smoothed particle
  hydrodynamics.
\newblock {\em Journal of computational physics}, 191(2):448--475, 2003.

\bibitem{craven1978smoothing}
P.~Craven and G.~Wahba.
\newblock Smoothing noisy data with spline functions.
\newblock {\em Numerische mathematik}, 31(4):377--403, 1978.

\bibitem{cunningham2010manipulation}
A.~L. Cunningham, H.~Donaghy, A.~N. Harman, M.~Kim, and S.~G. Turville.
\newblock Manipulation of dendritic cell function by viruses.
\newblock {\em Current opinion in microbiology}, 13(4):524--529, 2010.

\bibitem{dehghan2004weighted}
M.~Dehghan.
\newblock Weighted finite difference techniques for the one-dimensional
  advection--diffusion equation.
\newblock {\em Applied Mathematics and Computation}, 147(2):307--319, 2004.

\bibitem{dehghan2006finite}
M.~Dehghan.
\newblock Finite difference procedures for solving a problem arising in
  modeling and design of certain optoelectronic devices.
\newblock {\em Mathematics and Computers in Simulation}, 71(1):16--30, 2006.

\bibitem{dougan2012numerical}
N.~Do{\u{g}}an.
\newblock Numerical treatment of the model for hiv infection of cd4+ t cells by
  using multistep laplace adomian decomposition method.
\newblock {\em Discrete Dynamics in Nature and Society}, 2012, 2012.

\bibitem{el2017spectral}
G.~I. El-Baghdady, M.~Abbas, M.~El-Azab, and R.~El-Ashwah.
\newblock The spectral collocation method for solving (hiv-1) via legendre
  polynomials.
\newblock {\em International Journal of Applied and Computational Mathematics},
  3(4):3333--3340, 2017.

\bibitem{el2011fractional}
M.~El-Shahed and A.~Alsaedi.
\newblock The fractional sirc model and influenza a.
\newblock {\em Mathematical problems in Engineering}, 2011, 2011.

\bibitem{esmaeilbeigi2014new}
M.~Esmaeilbeigi and M.~Hosseini.
\newblock A new approach based on the genetic algorithm for finding a good
  shape parameter in solving partial differential equations by kansa’s
  method.
\newblock {\em Applied Mathematics and Computation}, 249:419--428, 2014.

\bibitem{fasshauer2007meshfree}
G.~E. Fasshauer.
\newblock {\em Meshfree approximation methods with MATLAB}, volume~6.
\newblock World Scientific, 2007.

\bibitem{fasshauer2007choosing}
G.~E. Fasshauer and J.~G. Zhang.
\newblock On choosing “optimal” shape parameters for rbf approximation.
\newblock {\em Numerical Algorithms}, 45(1):345--368, 2007.

\bibitem{gandomani2016numerical}
M.~R. Gandomani and M.~T. Kajani.
\newblock Numerical solution of a fractional order model of hiv infection of
  cd4+ t cells using m{\"u}ntz-legendre polynomials.
\newblock {\em International Journal Bioautomation}, 20(2):193, 2016.

\bibitem{ghoreishi2011application}
M.~Ghoreishi, A.~M. Ismail, and A.~Alomari.
\newblock Application of the homotopy analysis method for solving a model for
  hiv infection of cd4+ t-cells.
\newblock {\em Mathematical and Computer Modelling}, 54(11-12):3007--3015,
  2011.

\bibitem{gingold1977smoothed}
R.~A. Gingold and J.~J. Monaghan.
\newblock Smoothed particle hydrodynamics: theory and application to
  non-spherical stars.
\newblock {\em Monthly notices of the royal astronomical society},
  181(3):375--389, 1977.

\bibitem{goldberg1991real}
D.~Goldberg.
\newblock Real-coded genetic algorithms, virtual alphabeths, and blocking
  complex systems.
\newblock 1991.

\bibitem{gonzalez2014fractional}
G.~Gonz{\'a}lez-Parra, A.~J. Arenas, and B.~M. Chen-Charpentier.
\newblock A fractional order epidemic model for the simulation of outbreaks of
  influenza a (h1n1).
\newblock {\em Mathematical methods in the Applied Sciences},
  37(15):2218--2226, 2014.

\bibitem{hardy1971multiquadric}
R.~L. Hardy.
\newblock Multiquadric equations of topography and other irregular surfaces.
\newblock {\em Journal of geophysical research}, 76(8):1905--1915, 1971.

\bibitem{hassani2022optimal}
H.~Hassani, Z.~Avazzadeh, J.~T. Machado, P.~Agarwal, and M.~Bakhtiar.
\newblock Optimal solution of a fractional hiv/aids epidemic mathematical
  model.
\newblock {\em Journal of Computational Biology}, 29(3):276--291, 2022.

\bibitem{hemami2020use}
M.~Hemami, J.~A. Rad, and K.~Parand.
\newblock The use of space-splitting rbf-fd technique to simulate the
  controlled synchronization of neural networks arising from brain activity
  modeling in epileptic seizures.
\newblock {\em Journal of Computational Science}, 42:101090, 2020.

\bibitem{hemami2021phase}
M.~Hemami, J.~A. Rad, and K.~Parand.
\newblock Phase distribution control of neural oscillator populations using
  local radial basis function meshfree technique with application in epileptic
  seizures: A numerical simulation approach.
\newblock {\em Communications in Nonlinear Science and Numerical Simulation},
  103:105961, 2021.

\bibitem{holland1992adaptation}
J.~H. Holland.
\newblock {\em Adaptation in natural and artificial systems: an introductory
  analysis with applications to biology, control, and artificial intelligence}.
\newblock MIT press, 1992.

\bibitem{ibrahim2013new}
S.~Ibrahim and S.~Ismail.
\newblock A new modification of the differential transform method for a sirc
  influenza model.
\newblock {\em International Journal of Computer Applications}, 69(19), 2013.

\bibitem{johnson2012numerical}
C.~Johnson.
\newblock {\em Numerical solution of partial differential equations by the
  finite element method}.
\newblock Courier Corporation, 2012.

\bibitem{kansa1990multiquadrics}
E.~J. Kansa.
\newblock Multiquadrics—a scattered data approximation scheme with
  applications to computational fluid-dynamics—i surface approximations and
  partial derivative estimates.
\newblock {\em Computers \& Mathematics with applications}, 19(8-9):127--145,
  1990.

\bibitem{kermack1927contribution}
W.~O. Kermack and A.~G. McKendrick.
\newblock A contribution to the mathematical theory of epidemics.
\newblock {\em Proceedings of the royal society of london. Series A, Containing
  papers of a mathematical and physical character}, 115(772):700--721, 1927.

\bibitem{khader2014legendre}
M.~Khader and M.~M. Babatin.
\newblock Legendre spectral collocation method for solving fractional sirc
  model and in fluenza a.
\newblock {\em Journal of Computational Analysis \& Applications}, 17(2), 2014.

\bibitem{khader2014numerical}
M.~Khader, N.~Sweilam, A.~Mahdy, and N.~K. Moniem.
\newblock Numerical simulation for the fractional sirc model and influenza a.
\newblock {\em Applied Mathematics \& Information Sciences}, 8(3):1029, 2014.

\bibitem{khalili2022local}
A.~Khalili, V.~Ghanbari, and M.~Hemami.
\newblock A local scheme for numerical simulation of multi-dimensional dynamic
  quantum model: Application to decision-making.
\newblock {\em International Journal of Applied and Computational Mathematics},
  8(4):1--33, 2022.

\bibitem{khan2013efficient}
Y.~Khan, H.~Vazquez-Leal, and Q.~Wu.
\newblock An efficient iterated method for mathematical biology model.
\newblock {\em Neural Computing and Applications}, 23(3):677--682, 2013.

\bibitem{krysl1999element}
P.~Krysl and T.~Belytschko.
\newblock The element free galerkin method for dynamic propagation of arbitrary
  3-d cracks.
\newblock {\em International Journal for Numerical Methods in Engineering},
  44(6):767--800, 1999.

\bibitem{liu2005introduction}
G.-R. Liu and Y.-T. Gu.
\newblock {\em An introduction to meshfree methods and their programming}.
\newblock Springer Science \& Business Media, 2005.

\bibitem{longo2012harrison}
D.~L. Longo.
\newblock {\em Harrison: principios de medicina interna (18a}.
\newblock McGraw Hill Mexico, 2012.

\bibitem{merdan2007homotopy}
M.~Merdan.
\newblock Homotopy perturbation method for solving a model for hiv infection of
  cd4+ t cells.
\newblock {\em {\.I}stanbul Ticaret {\"U}niversitesi Fen Bilimleri Dergisi},
  6(12):39--52, 2007.

\bibitem{merdan2011approximate}
M.~Merdan, A.~Gokdogan, and V.~Erturk.
\newblock An approximate solution of a model for hiv infection of cd4+ t cells.
\newblock {\em Iranian Journal of Science and Technology (Sciences)},
  35(1):9--12, 2011.

\bibitem{moayeri2022npds}
M.~M. Moayeri, M.~Hemami, J.~A. Rad, and K.~Parand.
\newblock Npds toolbox: Neural population (de) synchronization toolbox for
  matlab.
\newblock {\em Neurocomputing}, 506:206--212, 2022.

\bibitem{oluwaseun2021block}
A.~Oluwaseun and O.~Zurni.
\newblock Block method for the solution of first order nonlinear odes and its
  application to hiv infection of cd4 t cells model.
\newblock 2021.

\bibitem{ongun2011laplace}
M.~Y. Ongun.
\newblock The laplace adomian decomposition method for solving a model for hiv
  infection of cd4+ t cells.
\newblock {\em Mathematical and Computer Modelling}, 53(5-6):597--603, 2011.

\bibitem{parand2021unsteady}
K.~Parand, S.~Hashemi-Shahraki, and M.~Hemami.
\newblock Unsteady flow of gas in a semi-infinite porous medium: a numerical
  investigation by using rbf-dqm.
\newblock {\em Indian Journal of Physics}, 95(10):2107--2114, 2021.

\bibitem{parand2017two}
K.~Parand, M.~Hemami, and S.~Hashemi-Shahraki.
\newblock Two meshfree numerical approaches for solving high-order singular
  emden--fowler type equations.
\newblock {\em International Journal of Applied and Computational Mathematics},
  3(1):521--546, 2017.

\bibitem{parand2018quasilinearization}
K.~Parand, Z.~Kalantari, and M.~Delkhosh.
\newblock Quasilinearization-lagrangian method to solve the hiv infection model
  of cd4+ t cells.
\newblock {\em SeMA Journal}, 75(2):271--283, 2018.

\bibitem{parand2018shifted}
K.~Parand, S.~Latifi, and M.~Moayeri.
\newblock Shifted lagrangian jacobi collocation scheme for numerical solution
  of a model of hiv infection.
\newblock {\em SeMA Journal}, 75(3):379--398, 2018.

\bibitem{parand2018pseudospectral}
K.~Parand, F.~Mirahmadian, and M.~Delkhosh.
\newblock The pseudospectral legendre method for solving the hiv infection
  model of cd4+ t cells.
\newblock {\em Nonlinear Studies}, 25(1), 2018.

\bibitem{parand2019numerical}
K.~Parand, M.~M. Moayeri, and S.~Latifi.
\newblock A numerical study on a model for hiv infection of cd4+ t-cells by
  shifted chebyshev polynomials.
\newblock {\em International Journal Bioautomation}, 23(2):163, 2019.

\bibitem{parand2013kansa}
K.~Parand and J.~Rad.
\newblock Kansa method for the solution of a parabolic equation with an unknown
  spacewise-dependent coefficient subject to an extra measurement.
\newblock {\em Computer Physics Communications}, 184(3):582--595, 2013.

\bibitem{perelson1989modeling}
A.~S. Perelson.
\newblock Modeling the interaction of the immune system with hiv.
\newblock In {\em Mathematical and statistical approaches to AIDS
  epidemiology}, pages 350--370. Springer, 1989.

\bibitem{perelson1999mathematical}
A.~S. Perelson and P.~W. Nelson.
\newblock Mathematical analysis of hiv-1 dynamics in vivo.
\newblock {\em SIAM review}, 41(1):3--44, 1999.

\bibitem{rad2015pricing}
J.~A. Rad, K.~Parand, and L.~V. Ballestra.
\newblock Pricing european and american options by radial basis point
  interpolation.
\newblock {\em Applied Mathematics and Computation}, 251:363--377, 2015.

\bibitem{rippa1999algorithm}
S.~Rippa.
\newblock An algorithm for selecting a good value for the parameter c in radial
  basis function interpolation.
\newblock {\em Advances in Computational Mathematics}, 11(2):193--210, 1999.

\bibitem{samui2020mathematical}
P.~Samui, J.~Mondal, and S.~Khajanchi.
\newblock A mathematical model for covid-19 transmission dynamics with a case
  study of india.
\newblock {\em Chaos, Solitons \& Fractals}, 140:110173, 2020.

\bibitem{sarra2005adaptive}
S.~A. Sarra.
\newblock Adaptive radial basis function methods for time dependent partial
  differential equations.
\newblock {\em Applied Numerical Mathematics}, 54(1):79--94, 2005.

\bibitem{sharan1997application}
M.~Sharan, E.~Kansa, and S.~Gupta.
\newblock Application of the multiquadric method for numerical solution of
  elliptic partial differential equations.
\newblock {\em Applied Mathematics and Computation}, 84(2-3):275--302, 1997.

\bibitem{takeda1994numerical}
H.~Takeda, S.~M. Miyama, and M.~Sekiya.
\newblock Numerical simulation of viscous flow by smoothed particle
  hydrodynamics.
\newblock {\em Progress of theoretical physics}, 92(5):939--960, 1994.

\bibitem{taylor1973numerical}
C.~Taylor and P.~Hood.
\newblock A numerical solution of the navier-stokes equations using the finite
  element technique.
\newblock {\em Computers \& Fluids}, 1(1):73--100, 1973.

\bibitem{thirumalai2021solution}
S.~Thirumalai, R.~Seshadri, and {\c{S}}.~Y{\"u}zba{\c{s}}{\i}.
\newblock On the solution of the human immunodeficiency virus (hiv) infection
  model using spectral collocation method.
\newblock {\em International Journal of Biomathematics}, 14(02):2050074, 2021.

\bibitem{umar2020stochastic}
M.~Umar, Z.~Sabir, F.~Amin, J.~L. Guirao, and M.~A.~Z. Raja.
\newblock Stochastic numerical technique for solving hiv infection model of
  cd4+ t cells.
\newblock {\em The European Physical Journal Plus}, 135(5):1--19, 2020.

\bibitem{umar2021neuro}
M.~Umar, Z.~Sabir, M.~A.~Z. Raja, J.~G. Aguilar, F.~Amin, and M.~Shoaib.
\newblock Neuro-swarm intelligent computing paradigm for nonlinear hiv
  infection model with cd4+ t-cells.
\newblock {\em Mathematics and Computers in Simulation}, 188:241--253, 2021.

\bibitem{venkatesh2016new}
S.~Venkatesh, S.~Raja~Balachandar, S.~Ayyaswamy, and K.~Balasubramanian.
\newblock A new approach for solving a model for hiv infection of cd 4+ t-cells
  arising in mathematical chemistry using wavelets.
\newblock {\em Journal of Mathematical Chemistry}, 54(5):1072--1082, 2016.

\bibitem{vespignani2020modelling}
A.~Vespignani, H.~Tian, C.~Dye, J.~O. Lloyd-Smith, R.~M. Eggo, M.~Shrestha,
  S.~V. Scarpino, B.~Gutierrez, M.~U. Kraemer, J.~Wu, et~al.
\newblock Modelling covid-19.
\newblock {\em Nature Reviews Physics}, 2(6):279--281, 2020.

\bibitem{webster1992evolution}
R.~G. Webster, W.~J. Bean, O.~T. Gorman, T.~M. Chambers, and Y.~Kawaoka.
\newblock Evolution and ecology of influenza a viruses.
\newblock {\em Microbiological reviews}, 56(1):152--179, 1992.

\bibitem{wendland2004scattered}
H.~Wendland.
\newblock {\em Scattered data approximation}, volume~17.
\newblock Cambridge university press, 2004.

\bibitem{xiang2012trigonometric}
S.~Xiang, K.-m. Wang, Y.-t. Ai, Y.-d. Sha, and H.~Shi.
\newblock Trigonometric variable shape parameter and exponent strategy for
  generalized multiquadric radial basis function approximation.
\newblock {\em Applied Mathematical Modelling}, 36(5):1931--1938, 2012.

\bibitem{yuzbacsi2012numerical}
{\c{S}}.~Y{\"u}zba{\c{s}}{\i}.
\newblock A numerical approach to solve the model for hiv infection of cd4+ t
  cells.
\newblock {\em Applied Mathematical Modelling}, 36(12):5876--5890, 2012.

\bibitem{zeb2013analytic}
A.~Zeb, G.~Zaman, M.~I. Chohan, S.~Momani, and V.~S. Ert{\"u}rk.
\newblock Analytic numeric solution for sirc epidemic model in fractional
  order.
\newblock {\em Asian Journal of Mathematics and Applications}, 2013, 2013.

\end{thebibliography}

\end{document}